\documentclass[sigconf, nonacm=true, anonymous=false]{acmart}

\usepackage{booktabs} 
\usepackage{graphicx}
\usepackage{subfigure}
\usepackage{algorithm}
\usepackage[noend]{algpseudocode}
\usepackage{xspace}
\usepackage{balance}
\graphicspath {{Figures/}}

\newcommand{\paratitle}[1]{\vspace{1.5ex}\noindent \textbf{#1}}
\newcommand{\ie}{\emph{i.e.,}\xspace}
\newcommand{\eg}{\emph{e.g.,}\xspace}

\setcopyright{none}

\begin{document}

\title{DP-GCN: Node Classification Based on Both Connectivity and Topology Structure Convolutions for Risky Seller Detection}

\author{Chen Zhe}
\affiliation{%
  \institution{Innovation Lab, PayPal}
  \country{Singapore}
}
\email{zchen1@paypal.com}

\author{Aixin Sun}
\affiliation{%
  \institution{Nanyang Technological University}
  \streetaddress{Nanyang Avenue}
  \country{Singapore}
}
\email{axsun@ntu.edu.sg}

\begin{abstract}
A payment network contains transactions between sellers and buyers. Detecting risky~(or bad) sellers on such a payment network is crucial to payment service providers for risk management and legal compliance. In this research, we formulate this task as a \textit{node classification} task. Specifically, we aim to predict a label for each seller in a payment network, by analysing its properties and/or interactions. 
Nodes residing in different parts of a payment network can have similar local topology structures. Such local topology structures  reveal sellers' business roles, \eg supplier, drop-shipper, or retailer. We note that many existing solutions for graph-based node classification only consider node connectivity but not the similarity between node's local topology structure. 
Motivated by business need, we present a \textit{dual-path graph convolution network}, named DP-GCN, for node classification. DP-GCN considers both \textit{node connectivity} and \textit{topology structure similarity}. The proposed model consists of three main modules: (i) a C-GCN module to capture connectivity relationships between nodes, (ii) a T-GCN module to capture topology structure similarities between nodes, and (iii) a multi-head self-attention module to align both properties. We evaluate DP-GCN on seven benchmark datasets against diverse baselines. We also provide a case study of running DP-GCN on three large-scale payment networks from PayPal, one of the leading payment service providers. Experimental results demonstrate DP-GCN's effectiveness and practicability.
\end{abstract}

\maketitle

\section{Introduction}  
\label{sec:intro}

Node classification is to infer labels of unclassified nodes in a network by exploiting node properties~\cite{jian2018toward}. Graph-based methods, especially graph convolution networks, have shown great success in various node classification applications. Example applications include payment network risk detection~\cite{lv2019auto,mcglohon2009snare}, social network analysis, anomaly detection~\cite{zheng2019addgraph}, and classification of molecules~\cite{ying2018hierarchical}.

One limitation of existing graph convolution based solutions is that they only model connectivity between nodes, but not other types of node properties, \eg nodes' local topology structures. The main reason is that convolution process in graph convolution networks only propagates information along edges.  Nodes residing at different parts of a graph may have similar local structural contexts, \eg star topology, ring topology. We refer such local structural identity of nodes their \textbf{local topology structures}. Many real-world applications require to take nodes' local topology structure into consideration. Take risky seller detection on payment network as an example. A large payment network typically contains millions of sellers and transactions. The risky sellers are those that result in revenue loss, due to different reasons \eg fraud, bankrupt, bad suppliers. In order to detect such risky sellers in advance, we need to consider sellers' connectivity relationships (\eg interactions with other sellers, buyers, and suppliers), as well as their topology structural similarities (\eg similarities on business models as supplier, drop-shipper, or retailer). We argue that existing graph network architectures are limited in modeling both properties. 

Motivated by this real-world business need, in this paper, we propose a \textbf{D}ual-\textbf{P}ath \textbf{G}raph \textbf{C}onvolution Attention \textbf{N}etwork architecture, named \textbf{DP-GCN}, for node classification. DP-GCN combines both connectivity and topology structure properties, in an efficient and effective manner. Specifically, DP-GCN consists of a C-GCN module, a T-GCN module, and a multi-head self-attention module. The C-GCN module captures the connectivity relationships between nodes; the T-GCN module captures the topology structure similarities between nodes; and the multi-head self-attention module aligns both properties. To model topology structure efficiently, we propose \textit{role-based convolution}, which extracts node topology structures into high-level \textit{latent roles} for convolution. This strategy significantly reduces the computation cost for each node, but well retains all convolution paths between nodes that are similar in terms of topology structures. 

In experiments, we evaluate DP-GCN on seven benchmark datasets. DP-GCN outperforms ten state-of-the-art baselines and achieves significant performance improvements on all datasets. We also conduct a careful ablation study and an embedding visualization to show that all components of DP-GCN are important for achieving the best performance. As a case study, we also evaluate DP-GCN on three large-scale payment networks from PayPal, to detect real risky sellers. This case study is challenging due to the size of networks, the diversity of risky seller types, and the high imbalance ratio. As expected, only a very small number of sellers are risky sellers. To summarize, we have made three contributions in this work.
\begin{itemize}
\item{We propose DP-GCN, a dual-path graph convolution attention network, for effective node classification, based on both node connectivity and topology structure similarity.}
\item{We propose role-based convolution, a novel and generalizable model, to make convolution process applicable to additional node properties aside connectivity.}
\item{We evaluate DP-GCN on seven benchmark datasets and also provide a case study on large-scale real-world datasets. Together with ablation study and embedding analysis, we show DP-GCN's effectiveness and practicality.}
\end{itemize}

\section{Related Work}   
\label{sec:related}

Our research is related to studies in four areas: node connectivity embedding, node topology structure embedding, graph neural network, and graph-based fraud detection.

\paratitle{Node Connectivity Embedding.}
Recent node embedding methods for graphs are largely based on the skip-gram model~\cite{mikolov2013efficient,cheng2006n}. In particular, Deepwalk~\cite{perozzi2014deepwalk} follows this approach to embed nodes to preserve nodes' neighborhood relationships through random walk. Node2vec~\cite{grover2016node2vec} optimizes this approach by adopting breadth-first and depth-first search in the random walk process. 
These methods become increasingly popular in both industry and research communities, as they outperform many existing methods with similar objectives. 
Due to the characteristics of random walk, those methods only focus on the connectivity relationship between nodes~\cite{tang2015line,perozzi2014deepwalk,grover2016node2vec}; hence these models do not consider roles of nodes based on their local topology structures in networks.

\paratitle{Node Topology Structure Embedding.}
Recently, many methods are proposed to capture the topology structure of nodes in embedding.
Struc2vec~\cite{ribeiro2017struc2vec} constructs a multi-layer weighted graph that encodes the topology structure similarity between nodes. If two nodes have similar degree and their neighbors also have similar degrees, then these two nodes are similar in topology structure. Graphwave~\cite{donnat2018learning} captures the topology structure of nodes by leveraging spectral graph wavelet diffusion patterns. 
RoIX~\cite{henderson2012rolx} extracts topology structure related features for nodes, then performs soft-clustering to group nodes based on those features.
DeepGL~\cite{rossi2018deep} constructs role features by local graphlet decomposition. It learns graph functions where each represents a composition of relational operators that are applied to a graphlet/motif feature. Hone~\cite{rossi2018hone} describes a framework based on weighted k-step motif graphs to learn  low-dimensional topology structure embeddings. Role2vec~\cite{ahmed2019role2vec} derives topology structure features by computing motif patterns of nodes. It then computes random walk transition probabilities based on topology structure similarities, and embeds nodes using  feature-based random walk. Recently, Riwalk~\cite{ma2019riwalk} is proposed to model node topology structure by its degree and relative position. It can better integrate graph kernels with node embedding methods to leverage recent advancements in node embedding.

\paratitle{Graph Neural Networks.}
With  advancement of deep learning, graph neural networks, especially graph convolution networks (GCN) have gained significant research interests~\cite{kipf2016semi,gao2018large,hamilton2017inductive,schlichtkrull2018modeling,velivckovic2017graph,liu2018heterogeneous}. For the task of node classification,  state-of-the-art solutions are dominated by GCN models. Different from skip-gram, GCN learns node representation by aggregating node's local neighbourhood through functions. GraphSage~\cite{hamilton2017inductive} extends the vanilla GCN by using neighborhood sampling during the aggregation process. It introduces several aggregation functions such as "Mean Pooling", "Max Pooling" and "LSTM function". GCNII~\cite{chen2020simple} is the deep graph neural network. Besides, GAT~\cite{ velivckovic2017graph} introduces self-attention mechanism to graph neural network. The self-attention measures the importance of different nodes in neighborhood when performing aggregation. DEMO-Net~\cite{wu2019net} models node degree specifically. It introduces degree-specific mapping function for feature aggregation in the convolution.

There are also studies on dual-path convolution architecture. The work from~\cite{zhuang2018dual} builds a PPMI matrix as the second path adjacency matrix to model graph global consistency. The authors in~\cite{zheng2020dual} use a dual-path architecture to embed image and text. It is a non-graph model but leverages dual-path to capture the relationship between image convolution and text convolution. DC-GCN~\cite{huang2020aligned} is a recent work for image question answering. It uses dual-path architecture to combine image and text for better answer prediction. AM-GCN~\cite{wang2020gcn} models node features with K-nearest neighbour (KNN) graph and propagates node features through both feature space and connectivity space simultaneously to better fusing node features and connectivity. 

Different from all existing dual-path studies, our model considers both node connectivity and node topology structure similarities. In particular, the role-based topology convolution model in the T-GCN module of DP-GCN is a novel convolution model. This new model is  generalizable and can be used to model other types of node properties. Moreover, the dual-path architecture designed in DP-GCN enables the two paths to be interactive, through a self-attention mechanism.

\paratitle{Graph-based Fraud Detection.}
Graph-based fraud detection usually focuses on analyzing graph connectivity and communities~\cite{pourhabibi2020fraud}. The system in~\cite{weng2018online} adapts HITS algorithm to detect fraudsters by propagating fraud score through graph. CatchSync~\cite{jiang2014catchsync} detects fraud by analyzing node properties~(\eg degree, HITS score, edge betweenness). The work~\cite{bindu2018discovering,manjunatha2018brnads,wang2018graph} detect fraud by analyzing fraudulent communities. GCNEXT~\cite{kudo2020gcnext} is a detection model using expanded balance theory and GCN. In summary, those methods assume that fraudsters in network are usually connected in communities. Thus all methods are able to detect fraudsters by exploring nodes' connectivity relationships.
In our study, through extensive experiments, we show that combining both graph connectivity and topology structure is important and effective for graph fraud~(risky seller) detection, and also node classification in general. Our work suggests a promising and practical direction for related applications.

\section{DP-GCN Architecture Overview}
\label{sec:framework}

\begin{figure*}[t]
    \centering
    \includegraphics[clip, trim=0cm 0cm 0cm 0cm, width=0.71\linewidth]{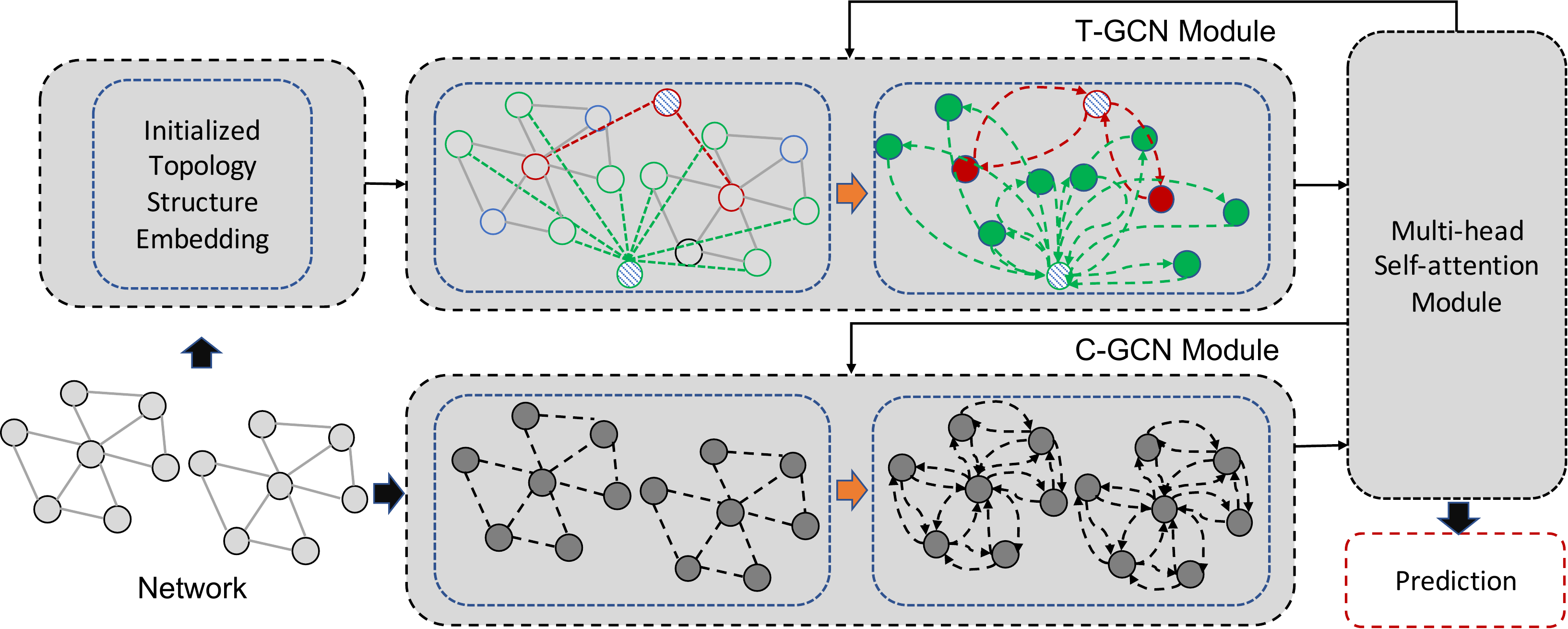}
\caption{Dual-path convolution with T-GCN and G-GCN (Best viewed in color). The shaded nodes in T-GCN are latent topology roles, constructed on top of existing nodes. The unified embedding produced by multi-head attention are shared as input for the next layer C-GCN and T-GCN. Thus both convolutions are interactively optimized towards the common classification objective.}
\label{fig:Framework}
\vspace{-3.5mm}
\end{figure*}

A graph is represented as $G = (V, E)$, where $V$ denotes the set of nodes and $E$ denotes the set of edges. The graph connectivity information is represented by a $|V|$ by $|V|$ adjacency matrix $A_c$, where $A_c[v_i][v_j] = 1$ if $v_i$ and $v_j$ are connected. Further, each node can be associated with node features. We use $F \in \mathbb{R}^{|V|\times{d}}$ to denote the feature matrix, where $d$ is the feature dimension. Usually, node features are specific to each individual node, \eg  demographic features of sellers in a payment network. Hence, node features are usually not related to the graph connectivity. In many related studies on graph convolution network, their solutions also support a non-feature option. In this case, an Identity matrix or Random matrix is used as feature matrix $F$~\cite{park2019exploiting}. In our problem setting, we take both node features $F$ and graph $G$ as input, and predict the label for each node.

The overall architecture of  DP-GCN is illustrated in Figure~\ref{fig:Framework}. We build our DP-GCN on top of the graph convolution framework, and its classification process is summarized as follows. Node representations are first initialized with the input features. We model node connectivity by the  C-GCN module and model topology structure similarity by the T-GCN module, through a dual-path convolution process. The outputs from both modules \ie C-GCN and T-GCN, are aligned by the multi-head self-attention module. This module automatically adjusts the importance of outputs from C-GCN and T-GCN, and learns the unified embedding representations of nodes. The learnt unified embeddings are then fed to the next layer of dual-path convolution, and eventually fed into the classification module for class label prediction. In the following sections, we detail all the modules in DP-GCN.

\section{Dual-Path Convolution}
\label{ssec:dp_convolution}
The dual-path convolution layer aims to learn node embeddings simultaneously for connectivity and for topology structure, through graph convolution in an end-to-end fashion. 

Reviewed in Section~\ref{sec:related}, related studies~\cite{zhuang2018dual,zheng2020dual,huang2020aligned} already show that dual-path convolution architecture is superior in learning unified node embedding. Our dual-path convolution consists of two interactive convolution modules: the C-GCN module and the T-GCN module. In our dual-path architecture, both modules take the unified embedding as input, thus the two modules will mutually reinforce each other, towards the common classification objective.

\subsection{The C-GCN Module}
\label{ssec:c-gcn}
The C-GCN module employs the standard graph convolution for modeling node connectivity relationship, which has been demonstrated to be effective in node representation learning.
The input to the connectivity convolution layer is the node feature matrix $F$, which represents the initial nodes' embeddings. All nodes connectivity information are stored in the adjacency matrix $A_c$, thus we directly apply convolution based on $A_c$. Specifically, we use the symmetric normalizing trick described in~\cite{kipf2016semi} to normalize $A_c$, to avoid changing the scale of feature vector:
\begin{equation}
\hat{A_c} = \hat{D}^{-\frac{1}{2}}({A_c}+I)\hat{D}^{-\frac{1}{2}}
\label{eqn:convolution}
\end{equation}
where  $\hat{D}$ is the diagonal degree matrix and $I$ is the Identity matrix. The Identity matrix is added to include each node's self-feature into convolution. Each convolution layer takes the input from previous layers and generates node level embedding $F^{(l+1)}$. The computation of the convolution is defined as:
\begin{equation}
F_{c}^{l+1} = \sigma(\hat{A_{c}}H^{l}W)
\end{equation}
In this equation,  $W$ is the weight matrix, $l$ denotes the $l-th$ convolution layer. $H$ denotes the unified embedding received from the multi-head self-attention module (see Figure~\ref{fig:Framework}), and the only exception is that $H^1=F$. The output of the C-GCN module $F_{c}^{l+1}$, will be fed into the multi-head self-attention module together with the output from the T-GCN module, to produce the updated unified embedding $H^{l+1}$.

Multi-layer convolutions enable the node to receive messages from a further neighbourhood~\cite{wu2020comprehensive}. As detailed above, our design is  different from a typical convolution layer.  The input to the successive convolution layers (\ie when $l>1$) in the C-GCN module is the unified node embedding $H^{l}$, rather than $F^{l}$ as in typical convolution. This design allows the two modules C-GCN and T-GCN to mutually reinforce each other while modeling different types of node properties.

\begin{figure}[t]
    \centering
    \includegraphics[clip, trim=0cm 0cm 0cm 0cm, width=1\linewidth]{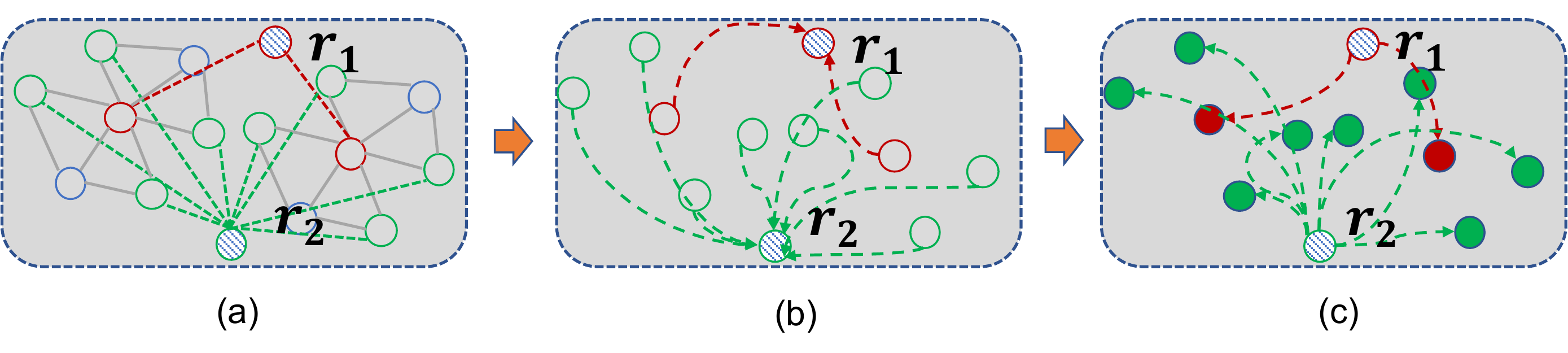}
\caption{(a) Topology roles are represented by shaded nodes ($r_1$, $r_2$). Member nodes connect to topology roles through belongingness edges (edges in green and red for the two topology roles). (b) First stage information propagation. Information is propagated from member nodes to topology roles. (c) Second stage information propagation. Information is propagated from topology roles to member nodes.}
\label{fig:Prop}
\vspace{-5mm}
\end{figure}

\subsection{The T-GCN Module}
\label{ssec:t-gcn}

The T-GCN module is designed to model the topology structure of nodes through graph convolution. The topology structure property is fine-tuned during the convolution, and finally integrated into the node embeddings. 

Before we present our role-based topology convolution, we discuss \textit{threshold-based convolution model}, a simple way to model topology structure by graph convolution. This simple model  generates "augmented paths" between nodes if their topology structure similarity is above a threshold. 
Those augmented paths create a new kind of neighborhoods for nodes that are similar in terms of local topology structures. In this way, graph convolution can aggregate nodes along the augmented paths, which naturally learn the node embedding based on topology structure similarity.

Specifically, a threshold-based convolution model is  designed in four steps: (i) leverage local topology structure embedding algorithms to generate node features representing topology structure, (ii) calculate pair-wise similarity between nodes by topology structure features,  (iii) create augmented paths between nodes if their topology structure similarity is above a predefined threshold, and (iv) perform graph convolution on augmented paths to fine-tune the node topology structure embedding. That is, we  derive a topology structure adjacency matrix $A_t$ to capture the augmented paths between nodes. The convolution in Equation~\ref{eqn:convolution} can be directly adopted on the augmented paths, by replacing $A_c$ with $A_t$. 

The time complexity for evaluating the pair-wise topology structure similarity between $n$ nodes is $O(n^{2})$. 
When a graph is large, it is impractical to evaluate and generate the augmented paths between every pair of nodes, for enterprise-scale applications. 
To address this limitation, we propose a novel \textbf{role-based convolution model} to perform topology structure convolution. 

\subsection{Topology Role Construction}

The main idea of role-based topology convolution is to summarize node topology structure with "latent topology roles". A \textbf{topology role} is a high-level representation of underlying topology structure identities of nodes, similar to the centroid of a cluster of nodes. 

To initialize topology roles, we first obtain topology structure features for each node. These features can be generated by using node topology structure embedding models (see Section~\ref{sec:related}).  Then we perform clustering or classification on topology structure features to discover the latent topology roles. Topology roles discovered by clustering preserve the original graph properties, while topology roles discovered by  classification can be directly mapped into the task-specific space (\ie node classification). 

Once topology roles are discovered, we create dummy nodes on the graph to represent these topology roles. Then, \textbf{belongingness edges} between topology roles and their member nodes are added to retain their topology structure similarities. If two nodes are similar by their topology structure features, then the two nodes connect to the same topology role through belongingness edges. 

In our implementation, we evaluate a few ways to generate topology structure features, \eg Struc2vec~\cite{ribeiro2017struc2vec}, Graphwave~\cite{donnat2018learning}, Role2vec~\cite{ahmed2019role2vec} and Riwalk~\cite{ma2019riwalk}. Based on the features, we discover the initial latent topology roles with $K$-means or regression classifier

\subsection{Topology Role Convolution}
After topology roles are identified, the convolution process is designed to extract the topology roles' representations, and to fine-tune their members' embeddings, simultaneously. The convolution consists of information propagation in two stages, as illustrated in Figure~\ref{fig:Prop}.

In the first stage,  information is propagated from member nodes to  topology roles, through belongingness edges. This process learns or updates the embedding representations for topology roles. The convolution on each topology role $r_i\in{R}$ is expressed by:
\begin{equation}
r_{i}^{l+1} = \sigma \left({\frac{1}{|{m_r({i})}|}}{\cdot}W\sum_{i\in{m_r({i})}}{{h^{l}_i}}\right)
\end{equation}
where $l$ denotes the $l-th$ convolution layer, $W$ is the weight parameter, and $m_r({i})$ denotes the member nodes of role $i$. Similarly, the input to the successive topology role convolution layers ($l>1$) are the unified node embedding $h\in{H}$.

In the second stage, the information is back-propagated from topology roles to their member nodes. This process fine-tunes the member nodes' embeddings from their topology roles.  Here, we assume that the underlying topology roles in the graph are non-overlapping. Thus we use a sharing scheme for this back-propagation. That is, nodes inherit the topology structure embeddings directly from their connected roles.

In implementation, the overall topology role convolution can be transformed into corresponding matrix multiplications as follows:
\begin{equation}
R^{l+1} = \sigma(\hat{A_{t}}H^{l}W)
\end{equation}
where $R^{l+1}$ is the embedding matrix of topology roles and $\hat{A_{t}}$ is the $|R| \times |N|$ normalized membership matrix for topology roles. 
Then, we compute the topology embedding for each member node by
\begin{equation}
F_{t}^{l+1} = \hat{A_{t}^{T}}R^{l+1}
\end{equation}
where $(\cdot)^T$ denotes  transpose. Note that, we assume topology roles are non-overlapping; hence the topology role embedding is shared among all its members. 

Role-based topology convolution is an effective way to model topology structure similarity between nodes. The strategy of using topology roles preserves the information propagation paths for all nodes that are similar in topology structures. As the convolution follows belongingness edges and each node has only one belongingness edge to its corresponding topology role, the complexity for the convolution is $O(m+n)$ for a graph with $n$ nodes and $m$  topology roles. Note that, the total number of topology roles is bounded by $O(n)$. In the worst case, each node is modeled by an individual topology role. But in this case, the topology structure information becomes trivial as there is no pair of nodes having similar topology structures.  



With the T-GCN module, the proposed dual-path convolution framework is generalizable to model other types of node properties besides topology structure. It is also extendable to a multi-path convolution framework when there are more properties to be modeled.

\section{Multi-Head Graph Self-Attention}
\label{ssec:mhattention}
Based on previous studies~\cite{vaswani2017attention, velivckovic2017graph}, we use a modified multi-head self-attention mechanism to enhance the correlation,  and to adjust the importance, between connectivity relationship and topology structure similarities, in each layer of dual-path convolution. For simplicity, we denote node connectivity property by $c$ and topology structure similarity property by $t$.  We obtain the attention coefficients for each property $j\in \{c, t\}$ of node $i$ as follows: 
\begin{equation}
\begin{split}
e^{l+1}_{ij} &= LeakyReLU \left(\alpha\left [Wh_i^l\Vert{W{f^{l+1}_{ij}}}\right]\right), j\in{\{c,t\}}\\
\end{split}
\end{equation}
Again, $l$ denotes the $l-th$ convolution layer; $h_i$ is the unified embedding and $f_{ij}$ is the output from either C-GCN or T-GCN, \ie $j\in \{c,t\}$. $\alpha$ is the weight of the transformation network while $LeakyReLU$ is the nonlinearity function. $\Vert$ denotes concatenation.   Following~\cite{velivckovic2017graph}, a shared linear transformation $W$ is applied to the input features $h_i$ and $f_{ij}$ to enhance the expressive power.
As $e^{l+1}_{ij}$ indicates the importance of the information to node $i$, the attention weight can be naturally calculated by SoftMax function:

\begin{equation}
a^{l+1}_{ij} = SoftMax_{ij}\left(e^{l+1}_{ij}\right) = \frac{\exp\left({e^{l+1}_{ij}}\right)}{\sum_{k\in{c,t}}{\exp\left(e^{l+1}_{ik}\right)}}
\end{equation}
Thus the unified embedding is obtained by:
\begin{equation}
h^{l+1}_i = \sigma \left(\sum_{{j}\in{\{c,t\}}}a^{l+1}_{ij}Wf^{l+1}_{ij} \right)
\end{equation}

Further, we adapt multi-head attention mechanism to jointly extract different types of information
from multiple representation subspaces~\cite{voita2019analyzing,li2018multi}. Similar to ~\cite{velivckovic2017graph}, the $K$ head attentions are concatenated and we obtain the unified embedding as:

\begin{equation}
h^{l+1}_i = \Vert^{K}_{k=1}\sigma \left(\sum_{{j}\in{\{c,t\}}}a^{l+1,k}_{ij}Wf^{l+1,k}_{ij}\right)
\label{eqn:MHattention}
\end{equation}

\begin{figure}[t]
    \centering
    \includegraphics[clip, trim=0cm 0cm 0cm 0cm, width=0.55\linewidth]{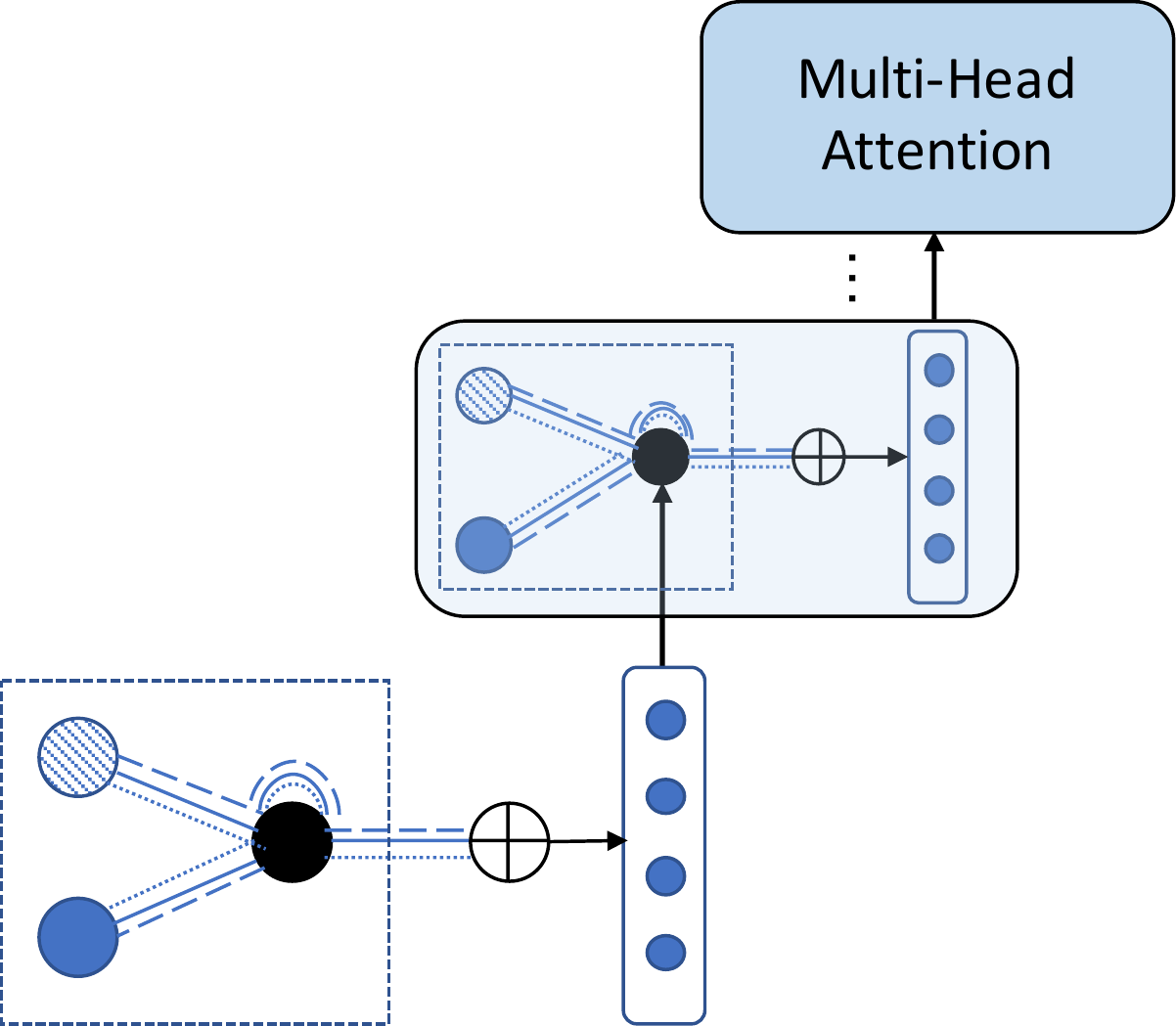}
\caption{Multi-head attentions in  different layers of convolution. The black node denotes the unified embedding from the previous layer. The blue and the shaded nodes are output embeddings from C-GCN and T-GCN respectively.}
\label{fig:MHAttention}
\end{figure}

Different convolution layers contain different hops of neighborhood information, thus we employ this attention on each layer of dual-path convolution. The attention mechanism on each convolution layer is illustrated in Figure~\ref{fig:MHAttention}. Following Equation~\ref{eqn:MHattention}, the attention on each layer takes the previously unified embedding into consideration, thus preserves the dependency information through layers. In our experiments, we employ single-head attention in the last dual-path convolution layer to generate unified embedding with fixed dimensions.

\section{Classification and Optimization}
\label{ssec:ag_construction}
Finally, the unified embeddings are fed into a classification network to do prediction. The  prediction layer is computed as follows:
\begin{equation}
\begin{split}
h^{\prime}_i &= Elu(h_i)\\
\hat{y_i} &= LogSoftMax(h^{\prime}_i)
\end{split}
\end{equation}

According to~\cite{zhu2020improving} and also our empirical evaluations, it is beneficial to add an additional normalization layer on top of the unified embedding for classification. With normalization, the predictions are expressed as: 
\begin{equation}
\begin{split}
h^{\prime}_i &= h^{\prime}_i/\Vert{h^{\prime}_i}\Vert_2\\
\hat{y_i} &= log\left (\frac{\exp({h^{\prime}_i})}{\sum{\exp({h^{\prime}_i})}}\right)
\end{split}
\label{eqn:pred}
\end{equation}

The network is trained by using standard multi-class negative log-likelihood loss in tandem with log-SoftMax in Equation~\ref{eqn:pred}~\cite{kanai2018sigsoftmax}. The negative log-likelihood loss is expressed as

\begin{equation}
L = -\sum_{i=1}^{n}\left[y_{i}\log{\hat{y_i}+(1-y_i)\log{(1-\hat{y_i})}}\right]
\end{equation}
where $y_{i}$ is the ground-truth label, and $\hat{y_i}$ is the output of the network.

\section{Experiments on Open Datasets}
In this section, we evaluate  DP-GCN classification model against ten state-of-the-art baselines on open networks. Through experiments, we also aim to find out  (i) whether  the T-GCN module can effectively utilize the topology structure information for classification, and (ii) whether the multi-head self-attention contributes to model performance. Lastly, we visualize the node representations learned from dual-path convolution. In Section~\ref{sec:casestudy}, we conduct a case study on PayPal payment networks. 

\subsection{Experiment Setting and Results}
We first evaluate our model on the task of node classification on five publicly available network datasets and two synthetic datasets, against baselines. All the five open network datasets come with ground-truth labels for validation.

\paratitle{Datasets.} We utilize five network datasets that take both connectivity and topology structure into consideration, for node classification. Among the five datasets, \textbf{Euro}, \textbf{Brazil} and \textbf{USA} are well known airline networks~\cite{ribeiro2017struc2vec,donnat2018learning}. Those networks encode direct flights between different airports. The label to each airport is its level of activity, which expects the topology structure equivalence due to hub-and-spoke structure existing in the datasets. The \textbf{BA} dataset is a labeled network for node classification research from~\cite{nr} and \textbf{Cora} is a labeled citation network. The relationship in Cora network focuses more on connectivity instead of topology structure. The two synthetic networks \textbf{Acm1} and \textbf{Acm2} are generated by~\cite{maekawa2019general}, and are included in our experiments for comprehensiveness. Data statistics are reported in Table~\ref{tab:dataset}.

\begin{table}[t]
  \small
    \caption{Statistics of four open and two synthetic networks}
  \label{tab:dataset}
  \begin{tabular}{c|rrrr}
    \toprule
    Dataset &\#Nodes& \#Edges & \#Classes & Type of data\\
    \midrule
    \texttt{Euro}& 399 & 5,995& 3 & Real data\\
    \texttt{Brazil}& 131& 1,074& 3& Real data\\
    \texttt{USA}& 1,190& 13,599& 3 & Real data\\
    \texttt{BA}& 804& 46,410& 5 & Real data\\ 
    \texttt{Cora}&2,708& 5,429& 7& Real data\\\midrule
    \texttt{Acm1}& 1,024& 22,008&6& Synthetic data\\
    \texttt{Acm2}& 4,096& 116,012&12& Synthetic data\\
  \bottomrule
  \end{tabular}
\vspace{-3mm}
\end{table}

\begin{table*}[t]
\caption{Evaluation on four open networks and two synthetic datasets; best results are in boldface and second best highlighted.}
  \footnotesize
  \label{tab:ResultExp1}
  \small{
  \begin{tabular}{c|ccccccc|ccccccc}
    \toprule
    & \multicolumn{7}{c|}{Accuracy} & \multicolumn{7}{c}{Fscore}\\
    \midrule
	Dataset & Euro& Brazil&USA& BA& Cora&Acm1& Acm2&  Euro& Brazil&USA& BA& Cora&Acm1& Acm2\\
    \midrule
    \texttt{Struc2vec}&0.588&0.630&0.534&0.230&0.602&0.263&0.112 &0.550&0.407&0.508&0.224&0.568&0.263&0.093\\
    \texttt{Graphwave}&0.438&0.444&0.576&0.168&0.210&0.136& -- &0.152&0.312&0.573&0.161&0.151&0.137&--\\
    \texttt{Role2vec}&0.363&0.519 &0.534&0.217&0.624&0.470&0.530&0.206&0.344&0.492&0.218&0.603&0.454&0.490\\
    \texttt{RiWalk}&0.575&\textbf{0.815}&0.601&0.199&0.421&0.215&0.115&0.482&0.558&0.365&0.185&0.363&0.187&0.100\\
    \texttt{Node2vec}&0.263&0.333&0.479&0.236&0.671&0.612&0.517&0.221&0.175&0.477&0.232&0.646&0.567&0.506\\
    \texttt{GCN}&0.400&0.481&0.462&0.224&0.891&0.675&0.500&0.368&0.300&0.454&0.073&0.878&0.667&0.466\\
    \texttt{GCNII}&0.522&0.500&0.577&0.237&\textbf{0.899}&0.746&0.414&0.460&0.494&0.523&0.230&\textbf{0.881}&0.732&0.330\\
    \texttt{GAT}&0.424&0.568&0.504&0.224&0.886&0.509&0.393&0.371&0.440&0.477&0.142&0.870&0.474&0.344\\
    \texttt{DEMO-Net}&0.525&0.593&0.588&0.273&0.710&0.502&0.545&0.545&0.529&0.557&\underline{0.250}&0.664&0.492&\underline{0.549}\\
    \texttt{AM-GCN}&0.375&0.482&0.332&0.217&0.884&\underline{0.761}&0.418&0.187&0.163&0.228&0.114&0.864&0.732&0.345\\
        \midrule
    \texttt{DP-GCN-SV}&0.613&\underline{0.741}&0.576&\textbf{0.286}&0.884&\textbf{0.766}&\underline{0.571}&0.582&\textbf{0.662}&0.562&\textbf{0.253}&0.871&\textbf{0.749}&0.527\\
        \texttt{DP-GCN-GW}&\underline{0.625}&0.704&\underline{0.605}&0.267&\underline{0.893}&0.746& -- &\underline{0.586}&0.488&\underline{0.589}&0.203&\underline{0.875}&0.718&--\\
    \texttt{DP-GCN-RV}&0.513&0.667&0.555&\underline{0.280}&0.887& 0.737&\textbf{0.573}&0.478&\underline{0.632}&0.540&0.235&0.873&0.716&\textbf{0.552}\\
    \texttt{DP-GCN-RiW}&\textbf{0.650}&0.704&\textbf{0.613}&0.255&0.886&0.756&0.560&\textbf{0.598}&0.594&\textbf{0.603}&0.173&0.873&\underline{0.739}&0.499\\
  \bottomrule
  \end{tabular}
  }
\end{table*}

\paratitle{Baseline Methods.} We consider both traditional and state-of-the-art graph-based methods as baselines. As different baselines use different strategies to study the graph, we ensure the baselines in our experiments demonstrate sufficient diversity. We include the following 10 baselines in our evaluation. \textbf{Struc2vec}~\cite{ribeiro2017struc2vec} is a widely used node topology structure embedding method. It encodes topology structure explored by random walk. \textbf{Graphwave}~\cite{donnat2018learning} learns node topology structure representation by diffusion patterns. \textbf{Role2vec}~\cite{ahmed2019role2vec} captures node topology structure similarity by analyzing node motif patterns. \textbf{Riwalk}~\cite{ma2019riwalk} models node topology structure by degree and relative position.
\textbf{Node2vec}~\cite{perozzi2014deepwalk} is an extension of the skip-gram embedding model by capturing breadth-first and depth-first search based neighborhood relationship. \textbf{GCN}~\cite{kipf2016semi} is a classic graph convolution neural network. \textbf{GCNII}~\cite{chen2020simple} is the deep graph neural network. \textbf{GAT}~\cite{velivckovic2017graph} adapts self-attention to measure the neighborhood importance when performing node convolution. \textbf{DEMO-Net} models node degree specifically and \textbf{AM-GCN} models node feature with KNN graph to better fusing its features and connectivity. To summarize, Struc2vec, Graphwave, Role2vec and Riwalk are node topology structure based methods,  Node2vec is connectivity based method. GCN, GCNII, GAT, DEMO-Net and AM-GCN are graph neural networks.

In our experiments, we split data for evaluation. Specifically,  $67\%$ of nodes are used for training and the remaining $33\%$ of nodes for testing (fully supervised). For GCN, GAT, DEMO-Net and DP-GCN, we use 2-layers of convolution and 120 hidden dimensions. For GCNII, we use 64-layers and 120 hidden dimensions. For AM-GCN, we use 1-layer of convolution and perform parameter search for $k$ (from 2 to 7) to build KNN graph. We use the Identity matrix as the initial feature input. For graph neural networks, we follow the standard strategy for training~\cite{kipf2016semi,velivckovic2017graph}. In specific, we build the entire graph for convolution, but only back-propagate the loss from the training nodes. For traditional methods, we implement the same classifier from DP-GCN and add additional dense layers for classification. We perform parameter search for hyper parameters (\eg number of hidden dimensions, number of layers) and adapt the optimal parameters for different datasets correspondingly. 

Recall that the T-GCN module requires initialization of topology roles.  We evaluate four different ways of topology roles initialization. Specifically, we initialize the topology roles in DP-GCN from Struc2vec (SV), Graphwave (GW), Role2vec (RV) and Riwalk (RiW) by k-means clustering ($k=100$) respectively. Accordingly, we have four variants of DP-GCN and we denote them by \textbf{DP-GCN-SV}, \textbf{DP-GCN-GW}, \textbf{DP-GCN-RV} and \textbf{DP-GCN-RiW} respectively.

\paratitle{Experimental Results.} We report both classification accuracy and Fscore for quantitative evaluation in Table~\ref{tab:ResultExp1}.\footnote{We did not manage to get results for Graphwave on Acm2, due to disconnectivity of the network.}

All four DP-GCN variants clearly outperform existing state-of-the-art models over a large margin. Specifically, the improvements are about $3\%$ to $11\%$. Especially on the Euro and Brazil networks, our methods have an average $> 8\%$ improvement in accuracy and Fscore over all other baselines. This result indicates the importance of combining both connectivity and topology structure information in node classification. We also observe that on Euro, Brazil, and USA datasets, the topology structure information is more important to the task, as the baselines focusing on topology structure information tend to have better performance. For BA, Cora and Acm1 networks, we observe the opposite. However, DP-GCN achieves consistently best results as our model utilizes both sides of information and uses self-attention to align them properly.  Overall, results of all variants DP-GCN-SV, DP-GCN-GW, DP-GCN-RV and DP-GCN-RiW  are comparable.

\begin{table}
  \caption{Ablation Analysis}
  \footnotesize
  \label{tab:Ablation}
  \begin{tabular}{l|ccccccc@{}}
    \toprule
    Module change& \multicolumn{7}{c}{Fscore}\\
    \midrule
	Dataset& Euro& Brazil& USA& BA&Cora&Acm1&Acm2\\
    \midrule
    \texttt{DP-GCN (Avg)}&\textbf{0.561}&\underline{0.594}&\textbf{0.573}&\textbf{0.216}&\textbf{0.873}&\textbf{0.730}&\textbf{0.526}\\ 
    \texttt{Remove C-GCN}&0.515&0.499&0.512&0.118&0.349&0.151&0.222\\
    \texttt{Remove T-GCN}&0.397&0.446&0.485&0.138&\underline{0.869}&0.668&0.435\\
    \texttt{Remove Att}&0.490& 0.458 &0.554&0.160&0.852&\underline{0.698}&0.495\\
    \texttt{1-Head Att}&\underline{0.540}&\textbf{0.603}&\underline{0.568}&0.133&0.868&0.693&\underline{0.523}\\
    \texttt{1-Layer GCN}&0.507&0.575&0.555&\underline{0.185}&0.854&\textbf{0.730}&0.457\\
  \bottomrule
  \end{tabular}

  \vspace{-3mm}
\end{table}

\subsection{Ablation Study}

We now conduct ablation studies to evaluate  the effectiveness of each module in the proposed DP-GCN model. Table~\ref{tab:Ablation} reports the results with different settings. As there are different ways to initialize the topology roles, we use the average Fscore of DP-GCN-SV/GW/RV/RiW when applicable. For instance, the first row in Table~\ref{tab:Ablation} is the average results of the four variants of DP-GCN.  

To evaluate the effectiveness of the component modules, we start with removing each individual module at a time, \ie C-GCN, T-GCN, and the multi-head attention module. When the multi-head attention module is removed, we combine the outputs from C-GCN and T-GCN unweighted. Alternatively, we also evaluate the setting of replacing the multi-head attention module by a single-head setting, or by using  a single convolution layer. 

We make four observations from Table~\ref{tab:Ablation}. First, both T-GCN and C-GCN modules play important roles in DP-GCN as the performance decreases significantly after removing either of two modules. Second, the attention mechanism is also an important component in DP-GCN. It is intuitive that different types of node information take different levels of importance in the prediction task. Combining multiple types of node information without carefully analyzing their importance may not be effective in the classification task, and sometimes may even introduce noise to the model. Third, tuning the number of heads for attention is not trivial. It helps the model to better learn the information alignment towards the task-specific objective from multiple representation subspaces. Finally, we observe that the C-GCN module does not contribute much to the Euro, Brazil, and USA datasets. This is due to the characteristics of the datasets. In those three datasets, the topology structure information is more important for the corresponding classification objective. However, on  BA, Cora, Acm1, and Acm2 datasets, the C-GCN module contributes more than the T-GCN module. Overall, this study suggests that all modules in the architecture are crucial for achieving the best classification results. If we combine the information modeled by both C-GCN and T-GCN modules with a proper attention mechanism, we are able to achieve the best classification results over all datasets.

\begin{figure}
\centering
\begin{subfigure}[Node representation from GCN\label{sfig:node-gcn}]{\includegraphics[trim=3cm 6cm 3cm 7cm, clip, width=0.75\columnwidth]{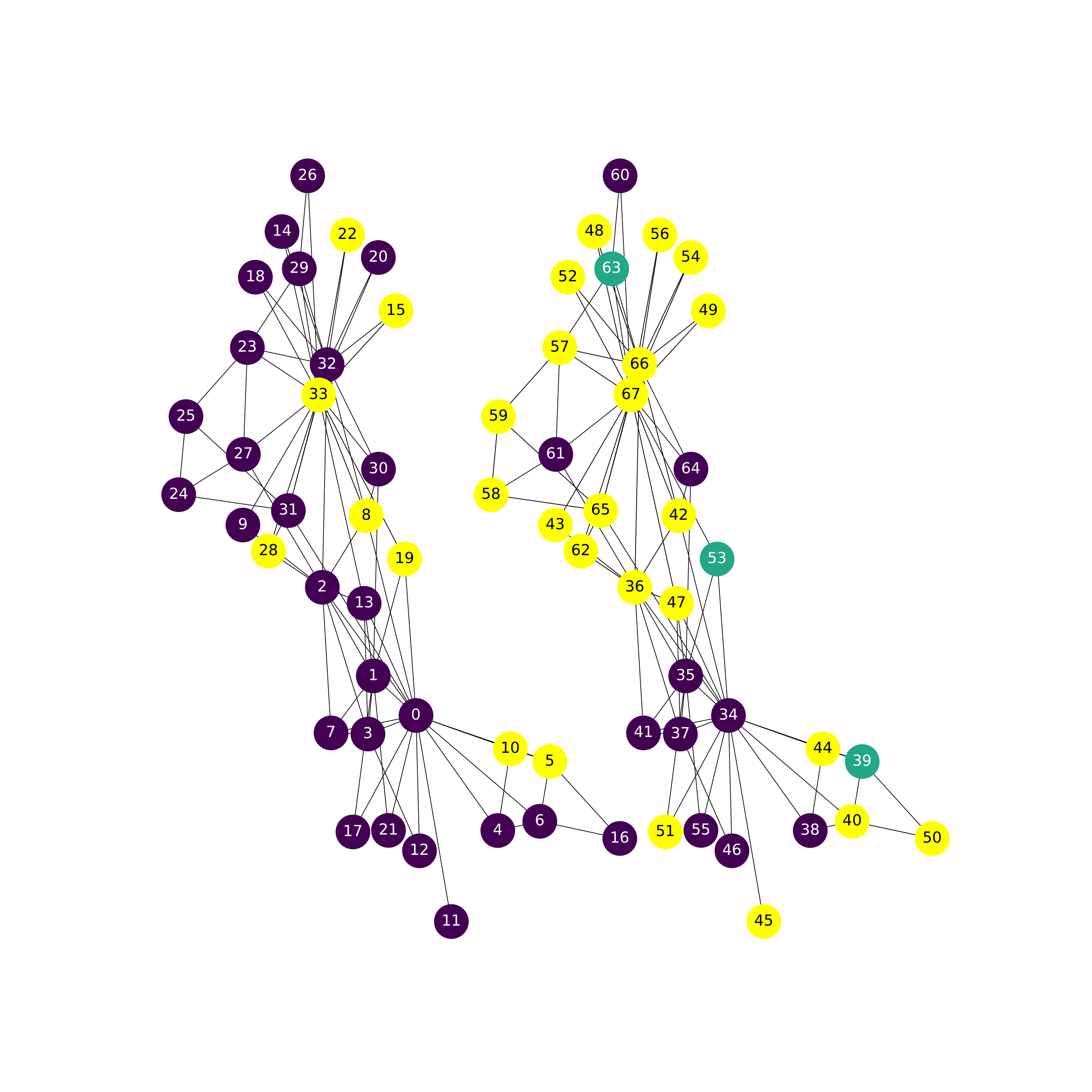}}
\end{subfigure}
\begin{subfigure}[Node representation from DP-GCN\label{sfig:node-dpgcn}]{\includegraphics[trim=3cm 6cm 3cm 7cm, clip, width=0.75\columnwidth]{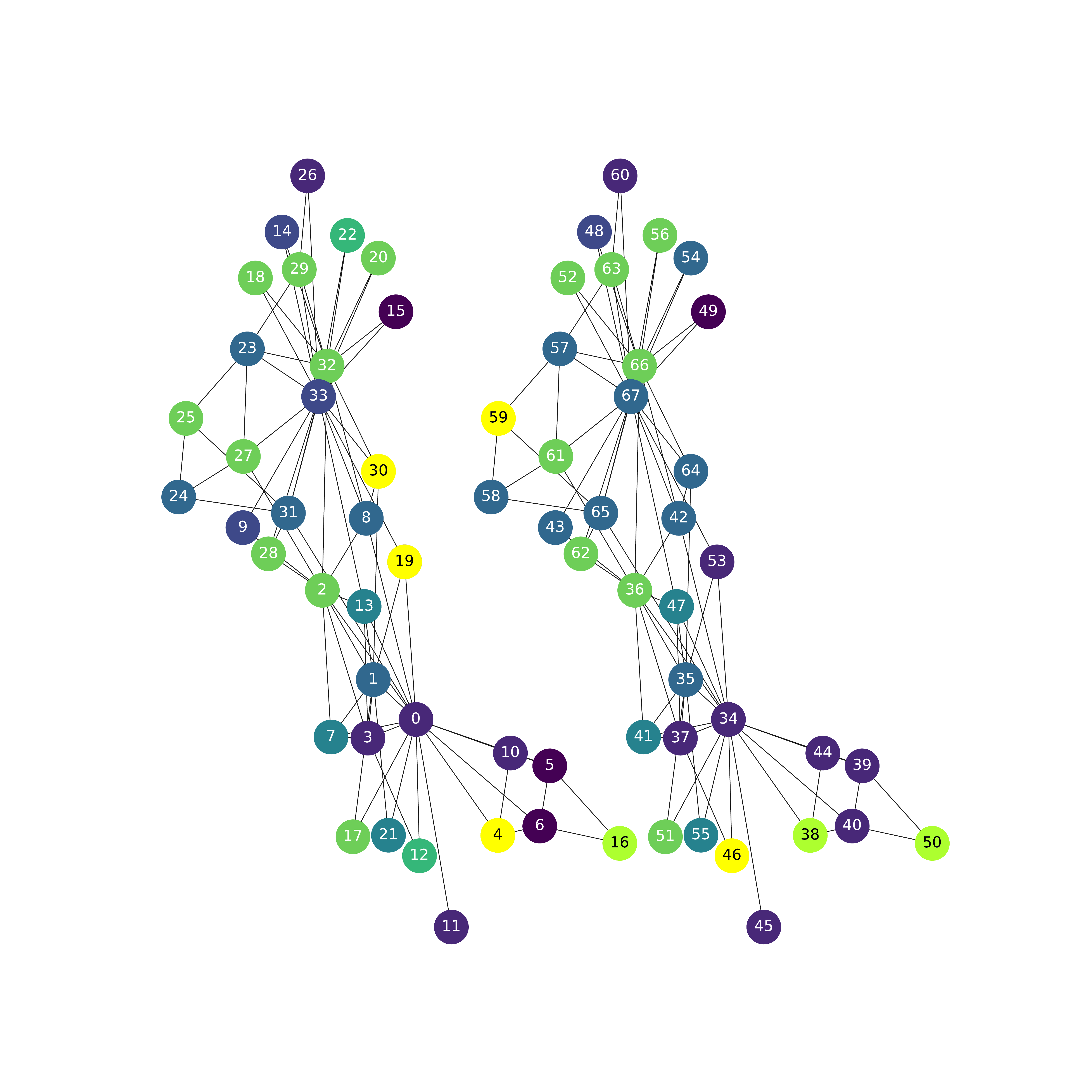}}
\end{subfigure}
\caption{Clusters of node representations from GCN, DP-GCN without training (Best viewed in color)}
\label{fig:representation}
\vspace{-4mm}
\end{figure}

\begin{table*}[t]
\footnotesize
 \caption{Risky seller detection on PayPal network datasets. Best results are in boldface and second best underlined. }
  \label{tab:CaseStudy}
  \begin{tabular}{c|ccc|ccc|ccc}
    \toprule
    Dataset& \multicolumn{3}{c|}{SG-Jan} &\multicolumn{3}{c|}{SG-Apr} &\multicolumn{3}{c}{SG-Jul}\\
    \midrule
	Model & Precision & Recall & Fscore& Precision & Recall & Fscore & Precision & Recall & Fscore\\
    \midrule
    \texttt{Role2vec}&0.523& 0.962 &0.678&0.545& 0.968 &0.698&0.523& \underline{0.961} &0.677\\
    \texttt{Riwalk}&0.526&0.956&0.679&0.531&\underline{0.969}&0.686&0.524&0.948&0.675\\
    \texttt{GCN}&0.518&0.854&0.645&0.521&0.884&0.655&0.520&0.904&0.660\\
    \texttt{GCNII}&\textbf{0.808}&0.507&0.623&\textbf{0.749}&0.502&0.601&\textbf{0.669}&0.501&0.573\\
    \texttt{DP-GCN-RV}&0.551&\underline{0.979}&\underline{0.705}&\underline{0.654}&0.902&\textbf{0.758}&\underline{0.561}&\underline{0.961}&\textbf{0.708}\\
    \texttt{DP-GCN-RiW}&\underline{0.567}&\textbf{0.987}&\textbf{0.720}&0.558&\textbf{0.985}&\underline{0.712}&0.526&\textbf{0.975}&\underline{0.683}\\
  \bottomrule
  
  \end{tabular}
  

\end{table*}
\subsection{Node Representation Visualization}

We now visualize the node representations learned by the dual-path convolution process in DP-GCN. This visualization may provide insights into the inner workings of our model. In specific, we visualize the node representations generated by  DP-GCN on the well known Zachary's Karate Club network~\cite{zachary1977information}. 

We follow~\cite{ribeiro2017struc2vec} to construct a mirrored network of the same club network. Then we obtain a unified network which consists of the original and the mirrored networks (\ie two copies of the same Karate Club network). In this unified network, each node is assigned a unique node id; so corresponding nodes in the original and mirrored networks will be assigned different node ids, as shown in Figure~\ref{fig:representation}. We build DP-GCN with randomly initialized weights on this unified network. To enable role convolution, we make a simple assumption that the same corresponding nodes in the original and the mirrored networks have the same topology structure (\ie modelled by the same topology role). Without any training, we input the adjacency matrix of the unified Zachary's karate club and the Identity matrix $X=I$ (\ie without any additional features) to the model. Then DP-GCN performs both connectivity and topology role convolution on all nodes in the unified network. The final node representations are obtained through a random weighted attention which combines both sides of information (as we do not have a task-specific loss function here to guide the attention). As a comparison, we generate node representations from the standard GCN through the same process. In this analysis, we set the dimension of node representations to be 10 and feed the learned node representations to a log-SoftMax function. The corresponding dimension obtained from the log-SoftMax is the node cluster label. Figure~\ref{fig:representation} shows nodes clusters in different colors.

Observe that GCN tends to group nodes based on connectivity relationships, thus nodes within the same local community have very close representations. The convolution from GCN does not take topology structure similarity into consideration. The representations generated from DP-GCN, on the other hand, consider both node connectivity and node topology structure similarity. As an example, the corresponding nodes \{18,29,32\} and \{52,63,66\} are densely connected in each of their own local communities. They have the same topology structure as they reside at the same positions in the original and the mirrored club networks. Thus they have been learned to have a close representation by DP-GCN, but not GCN. Results in Figure~\ref{fig:representation} verifies the effectiveness of using dual-path convolution to model these two types of information simultaneously in a network.

\section{Case Study}
\label{sec:casestudy}

As a case study, we evaluate the proposed DP-GCN model on payment networks from PayPal. The task is to detect risky sellers, \ie the sellers who result in revenue loss due to various reasons like fraud, bankrupt, and bad suppliers.

\subsection{PayPal Datasets}
We conduct experiments on three network datasets with ground-truth labels from PayPal: \textbf{SG-Jan}, \textbf{SG-Apr} and \textbf{SG-Jul}. Each network contains a subset of one-week payment activities of anonymous sellers (seller ids are encrypted) in January, April, and July 2020 respectively. The network includes all sellers located in Singapore region while the buyers are from all over the world. The sizes of the three datasets are comparable and each contains about 0.5 million nodes and 0.5 million edges.\footnote{Due to enterprise privacy, we mask the exact numbers.} The risky seller labels are provided by an internal department for various risky types. The data is imbalanced, due to the fact that only a very small number of sellers are risky sellers.

Due to memory constraints, we manage to compare our model DP-GCN with Role2vec, Riwalk, GCN, and GCNII. Again, Role2vec, Riwalk are node topology structure based methods. GCN and GCNII are the graph convolution networks. We use 2-layers and 120 hidden dimensions for DP-GCN and GCN. We use 64-layers and 120 hidden dimensions for GCNII. We perform parameter search on the over- and under-sampling ratio for the imbalanced labels and use the same ratio for all methods. For DP-GCN, we use topology roles generated from Role2vec and Riwalk embeddings, and use 1000-dimensional binary vectors as node initial features. For GCN and GCNII, we evaluate the performance by using either 1000-dimensional binary vectors or topology structure embeddings as node initial features, and then select the best one to represent its final performance. As the labels are much imbalanced, we use macro-averaged precision, recall, and Fscore for performance evaluation, instead of accuracy. Macro-averaged metrics are known to be more appropriate for imbalanced data as it avoids dominance of majority classes~\cite{zheng2004feature,choeikiwong2016improve}.

\subsection{Result Analysis}

Results of risky seller detection are reported in Table~\ref{tab:CaseStudy}. DP-GCN is consistently at the top across all the three datasets by Fscore. Specifically, the improvements are about $4\%$ to $6\%$. Role2vec and Riwalk are the best performing models among baselines, which suggests that modelling topology structure is important for identifying real risky sellers. We also observe that simply feeding topology structure embeddings as the initial features to graph convolution network leads to very marginal improvement. GCNII tends to provide conservative prediction but catches less risky seller. This result further emphasizes the importance of using dual-path convolution to model the different types of information in the graph neural network.
Our results on enterprise-scaled data suggests that combining both connectivity and topology structure information for risky seller detection is a promising direction on payment network. Our model is well aligned with the real-world enterprise needs.

\section{Conclusion}
In this paper, we propose an interactive dual-path graph convolution network to explore both node connectivity and topology structure information, for risky seller detection and also node classification in general. We propose a generalizable role-based convolution to explicitly model the topology structure similarity of nodes with latent topology roles. With the notion of topology role, the computational cost of convolution over node's local topology structure similarity becomes much lower.  Furthermore, we align node connectivity and topology structure information with multi-head graph self-attentions to reduce the gaps between the two types of information. Via visualization and qualitative analysis, we show that our proposed model achieves highly competitive results on both open datasets and PayPal's risky seller detection task. Our case study also shows that the proposed model can be applied to large networks with million nodes.



\bibliographystyle{ACM-Reference-Format}
\balance
\bibliography{RoleNodeClassification}

\end{document}